\documentclass[a4paper,fleqn,usenatbib,useAMS]{mnras}

\usepackage{mathptmx}

\usepackage[T1]{fontenc}
\usepackage{ae,aecompl}

\usepackage{graphicx,epstopdf}	
\usepackage{amsmath}	
\usepackage{amssymb}	
\usepackage{mathtools}

\newcommand{\der}{{\rm d}} 

\title[Gaps in low-viscosity discs]{Deep and wide gaps by super Earths in low-viscosity discs}

\author[S. Ginzburg and R. Sari]{
Sivan Ginzburg\thanks{E-mail: sivan.ginzburg@mail.huji.ac.il}
and Re'em Sari
\\
Racah Institute of Physics, The Hebrew University, Jerusalem 91904, Israel}

\date{Accepted XXX. Received YYY; in original form ZZZ}

\pubyear{2017}

\begin{document}
\label{firstpage}
\pagerange{\pageref{firstpage}--\pageref{lastpage}}
\maketitle

\begin{abstract}
Planets can open cavities (gaps) in the protoplanetary gaseous discs in which they are born by exerting gravitational torques. Viscosity counters these torques and limits the depletion of the gaps. We present a simple one-dimensional scheme to calculate the gas density profile inside gaps by balancing the gravitational and viscous torques. By generalizing the results of Goodman \& Rafikov (2001), our scheme properly accounts for the propagation of angular momentum by density waves. This method allows us to easily study low-viscosity discs, which are challenging for full hydrodynamical simulations. We complement our numerical integration by analytical equations for the gap's steady-state depth and width as a function of the planet's to star's mass ratio $\mu$, the gas disc's aspect ratio $h$, and its Shakura \& Sunyaev viscosity parameter $\alpha$. Specifically, we focus on low-mass planets ($\mu<\mu_{\rm th}\equiv h^3$) and identify a new low-viscosity regime, $\alpha<h(\mu/\mu_{\rm th})^5$, in which the classical analytical scaling relations are invalid. Equivalently, this low-viscosity regime applies to every gap that is depleted by more than a factor of $(\mu_{\rm th}/\mu)^3$ relative to the unperturbed density. We show that such gaps are significantly deeper and wider than previously thought, and consequently take a longer time to reach equilibrium.
\end{abstract}




\section{Introduction}\label{sec:introduction}

Young planets gravitationally interact with the gaseous discs that surround their host stars during their first few million years \cite[see][for a review]{KleyNelson2012}. Specifically, a planet on a circular orbit repels gas particles on adjacent circular orbits by increasing (decreasing) their angular momentum if they are outward (inward) to the planet \citep{LinPapaloizou79,GoldreichTremaine80}. Under certain circumstances \citep[e.g.][]{Rafikov2002_gap}, these repulsive torques enable planets to carve out cavities (gaps) in the gas disc.

Recent studies tried to calculate the shape (in particular, depth and width) of the gaps that planets open by numerically solving the hydrodynamic equations in two and three dimensional simulations \citep{DuffellMacFadyen2013,Fung2014,FungChiang2016}.
In viscous discs, the gap opening saturates when the planetary torque, which tends to deepen the cavity, is balanced by the viscous torque that fills it back \citep[e.g.][]{Fung2014}.
However, the physical origin of the viscosity is not well understood: molecular viscosity is far too low, while magnetorotational instability \citep[MRI,][]{BalbusHawley91} is complicated and uncertain \citep[e.g.][]{Bai2016}. This has lead several authors to suggest that protoplanetary discs have a very low intrinsic viscosity and that gas loses angular momentum (and thereby accretes onto the star) by planetary torques \citep{Larson89,GoodmanRafikov2001,SariGoldreich2004,FungChiang2017}.
Hydrodynamical simulations of such low-viscosity discs suffer from numerical viscosity, which is difficult to constrain \cite[e.g.][]{DuffellMacFadyen2013}. In addition, the time (number of planet orbits) to reach equilibrium increases with decreasing viscosity, lengthening the simulation runtime. For these reasons, there is an advantage for using analytical methods to calculate the gap profile.   

\citet{TanigawaIkoma07} provided an analytical solution that balances the planetary torque with viscosity and also accounts for the Rayleigh instability, which limits the curvature of the gas density profile inside the gap \citep[e.g.][]{YangMenou2010}. They assumed that the planetary torque deposits angular momentum locally. \citet{GoodmanRafikov2001}, on the other hand, showed that low-mass planets generate density waves that steepen until they shock, dissipate, and impart their angular momentum to the disc at a different location from the wave excitation region. \cite{Kanagawa2015} demonstrated that this non-local nature of the angular momentum deposition may significantly alter the gap's density profile. However, \citet{Kanagawa2015} modelled the non-locality with an arbitrarily chosen parameter that specifies the dissipation location. \citet{Duffell2015}, on the other hand, provided an analytical solution for the gap's shape that takes into account wave dissipation according to the \citet{GoodmanRafikov2001} mechanism. Despite its ability to reproduce relatively shallow gaps that are opened in viscous discs, the analytical solution of \citet{Duffell2015} suffers from two shortcomings, inherited from the \citet{GoodmanRafikov2001} wave-shocking condition. This condition is valid only for discs of roughly uniform density, and in addition, as \cite{Duffell2015} recognized, the relative contribution of the gap's periphery to the total torque increases as the density at the gap's centre decreases. For these two reasons, as we show in the current paper, the previous analytical solutions are invalid for deep gaps that are opened in low-viscosity discs.\footnote{Regardless of these shortcomings, it appears that the analytical model of \citet{Duffell2015} deviates from the numerical results for deep gaps, in Figures 2 and 4 in that paper, primarily due to a different reason. The planet's mass exceeds the thermal mass (defined in Section \ref{sec:non_local}) for high planet masses or for low disc aspect ratios, introducing highly non-linear effects \citep[see][and Section \ref{sec:thermal_mass}]{GoodmanRafikov2001}.}

Here, we generalize the results of \citet{GoodmanRafikov2001} and derive an analytical relation between the excitation and deposition locations of density waves travelling across a deep gap. We incorporate this relation into a simple one dimensional integration and obtain accurate gap density profiles that correctly account for the wave propagation. In addition, we provide a fully analytical approximation for the gap's profile that illuminates the results of the numerical integration.

The outline of the paper is as follows. In Section \ref{sec:local} we review the analytical solution for gap profiles assuming local angular momentum deposition. In Section \ref{sec:non_local} we extend the discussion to non-local deposition, describe our integration scheme, and provide an analytical description of the resulting density profiles.
In Section \ref{sec:results} we present analytical expressions for the gap's equilibrium depth and width and for the time it takes to reach this steady state. We then conduct a systematic numerical parameter survey and compare the results to the analytical prediction. We summarize our findings and discuss their implications in Section \ref{sec:summary}.

\section{Local deposition}\label{sec:local}

In this section we briefly review the analytical solution for a gap's density profile, under the simplifying assumption of local angular momentum deposition \citep[see also][]{LubowDAngelo06,TanigawaIkoma07,Kanagawa2015}.

We adopt a dimensionless notation by setting the gravitational constant $G$, the planet's orbital separation $a$, and the stellar mass ${\sim\rm M}_{\sun}$ to unity. We also omit order of unity coefficients in our order of magnitude analysis. In these units, the planet's mass is given by $\mu\ll 1$ and the disc's scale height (as well as the sound speed) is $h\ll 1$. We focus on planets several times the mass of Earth ${\rm M}_{\earth}$ that orbit solar-mass stars at a separation $a\sim 0.1\textrm{ AU}$ (equivalently, with periods of $\sim 10$ days). These short-period super Earths seem to be the most abundant type of planet found by the {\it Kepler} mission \citep[e.g.][]{WolfgangLopez2015}. For such a planet, $\mu\sim 10^{-5}$, while $h\approx 3\times 10^{-2}$ assuming a disc temperature of about $10^3\textrm{ K}$ \citep{ChiangLaughlin2013}.

We set the origin $x=0$ at the planet's location and calculate the density profile as a function of the radial distance $x\ll 1$ from the planet (from symmetry, we are only interested in $x>0$). 
The torque that the planet exerts on a gas annulus with a width $\der x$ at a distance $x>h$ was calculated by \citet{GoldreichTremaine80} and can also be obtained by the impulse approximation \citep{LinPapaloizou79,LubowIda2010}
\begin{equation}\label{eq:torque}
\der T=\frac{\Sigma\mu^2}{x^4}\der x,
\end{equation}  
with $\Sigma$ denoting the gas surface density. At $x<h$, the torque drastically decreases with decreasing $x$, such that the total torque is dominated by $x=h$ \citep{GoldreichTremaine80}. In equilibrium, this torque density is balanced by the differential viscous torque that acts on the annulus
\begin{equation}\label{eq:visc}
\frac{\der T}{\der x}=\nu\frac{\der \Sigma}{\der x}=\frac{\Sigma\mu^2}{x^4},
\end{equation}  
where we assume Keplerian rotation (justified in Section \ref{sec:rayleigh}) and a uniform kinematic viscosity $\nu\equiv\alpha h^2$, with $\alpha$ denoting the \cite{ShakuraSunyaev73} parameter.
By solving equation \eqref{eq:visc} we obtain the gas density profile inside the gap
\begin{equation}\label{eq:local_profile}
\ln\frac{\Sigma(x)}{\Sigma_\infty}=-\frac{\mu^2}{3\nu x^3},
\end{equation} 
where $\Sigma_\infty$ is the unperturbed gas density at infinity. We keep the coefficients in equations \eqref{eq:local_profile} and \eqref{eq:parabola} for a smooth attachment between the two, as explained in Section \ref{sec:rayleigh}.

\subsection{Rayleigh instability}\label{sec:rayleigh}

A differentially rotating disc must satisfy
\begin{equation}\label{eq:rayleigh}
\frac{\rm d}{{\rm d}r}\left(\Omega r^2\right)>0,
\end{equation}
with $r$ denoting the radius and $\Omega$ the angular velocity. Otherwise, the disc is unstable to the growth of angular-momentum conserving perturbations. It is easy to verify that a Keplerian disc ($\Omega_{\rm K}\propto r^{-3/2}$) is Rayleigh stable. However, protoplanetary gas discs are not exactly Keplerian, due to the gas pressure gradient
\begin{equation}\label{eq:sub_keplerian}
\Omega^2 r=\Omega_{\rm K}^2r+\frac{1}{\rho}\frac{{\rm d} P}{{\rm d}r},
\end{equation}
with $P$ and $\rho$ denoting the gas pressure and density, respectively. By inserting equation \eqref{eq:sub_keplerian} into equation \eqref{eq:rayleigh} we find the following stability criterion \citep[e.g.][]{YangMenou2010}
\begin{equation}\label{eq:stability}
\Omega_{\rm K}^2r^3+\frac{\rm d}{{\rm d} r}\left(\frac{r^3}{\rho}\frac{{\rm d}P}{{\rm d}r}\right)=\Omega_{\rm K}^2r^3+\frac{\rm d}{{\rm d} r}\left(\Omega_{\rm K}^2h^2r^3\frac{{\rm d}\ln\rho}{{\rm d} r}\right)>0,
\end{equation}
where the sound speed is given by $\Omega_{\rm K}h=\sqrt{{\rm d}P/{\rm d}\rho}$. It is again easy to verify that if the density and sound speed vary on a length scale of order $r$, and if $h\ll 1$, then the deviation of the gas disc from Keplerian rotation is small and it remains Rayleigh stable. Therefore, unperturbed gas discs, which are usually modelled as power laws in $r$ \citep{Weidenschilling77,Hayashi1981,ChiangLaughlin2013,Schlichting2014}, are stable. When a planet opens a gap in the disc, on the other hand, the resulting perturbation in density might be sharper, violating the stability criterion.
We expand equation \eqref{eq:stability} around $r=a=1$ (where $\Omega_{\rm K}=1$) for $x\equiv r-1\ll 1$
\begin{equation}\label{eq:curvature}
\frac{{\rm d}^2\ln\Sigma}{{\rm d} x^2}>-\frac{1}{h^2},
\end{equation}
where we use $\Sigma=\rho h$ and assume that the scale height $h$ (i.e. the disc's temperature) varies slowly, on a length scale of $r$ \citep[see][]{YangMenou2010,Kanagawa2015}. Equation \eqref{eq:curvature} limits the curvature of the gap profile.

The local-deposition profile, given by equation \eqref{eq:local_profile}, violates the stability criterion for $x<x_{\rm R}\equiv(4\mu^2/\alpha)^{1/5}>h$ (as explained below, the last inequality holds for gaps with significant depth). Therefore, for $h<x<x_{\rm R}$, $\ln\Sigma(x)$ is given by the marginally stable parabola according to equation \eqref{eq:curvature}
\begin{equation}\label{eq:parabola}
\ln\frac{\Sigma(x)}{\Sigma_\infty}=-\frac{x^2}{2h^2}+\frac{5x_{\rm R}x}{4h^2}-\frac{5x_{\rm R}^2}{6h^2}.
\end{equation}
Using equations \eqref{eq:local_profile} and \eqref{eq:sub_keplerian} it is easy to verify that the deviation from Keplerian rotation (and its derivatives) is small enough in the Rayleigh-stable part of the profile ($x>x_{\rm R}$) to justify its omission from the derivation of the profile there. 

\begin{figure}
	\includegraphics[width=\columnwidth]{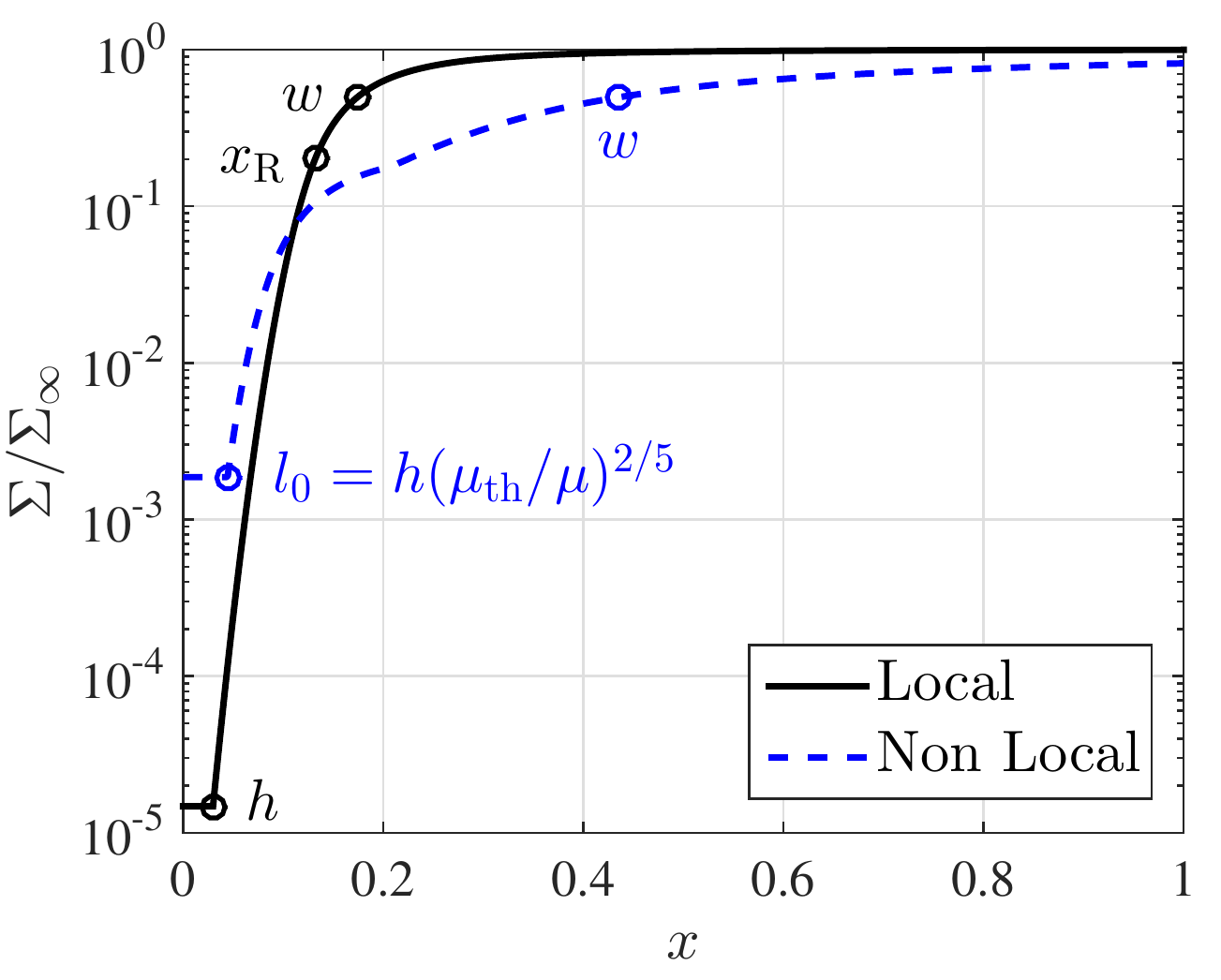}
	\caption{Equilibrium surface density profiles of gaps opened by a planet (at $x=0$) with a mass $\mu=10^{-5}$ in a disc with a scale height $h=3\times 10^{-2}$ and viscosity $\alpha=10^{-5}$. The profile under the assumption of local angular-momentum deposition (solid black line) is given by equation \eqref{eq:local_profile} for $x>x_{\rm R}$ and equation \eqref{eq:parabola} for $h<x<x_{\rm R}$ (see Section \ref{sec:rayleigh}). The profile that accounts for non-local deposition (dashed blue line) is given by an integration (see Section \ref{sec:integration}) of equations \eqref{eq:relation} and \eqref{eq:non_local_profile}. $h$, $x_{\rm R}$, and the width $w$ of the local profile are marked by black circles. $l_0$ (see Section \ref{sec:understand}) and the width of the non-local profile are marked by blue circles.}
	\label{fig:compare_local}
\end{figure}

An example of a density profile, under the assumption of local angular momentum deposition, is given in Fig. \ref{fig:compare_local}. Using equations \eqref{eq:local_profile} and \eqref{eq:parabola} we find that the depth of such a gap, i.e. the ratio of the density at the bottom of the gap $\Sigma_0$ to the unperturbed density $\Sigma_\infty$, is given by
\begin{equation}
\ln\frac{\Sigma_0}{\Sigma_\infty}=-\left(\frac{\mu^2}{\alpha h^5}\right)^{2/5},
\end{equation}
with comparable (up to coefficients of order unity) contributions from $x>x_{\rm R}$ and $h<x<x_{\rm R}$. The width of the gap $w$, defined here as the location for which $\Sigma(w)/\Sigma_\infty=1/2$, is found using equation \eqref{eq:local_profile}:
\begin{equation}\label{eq:width_local}
w\simeq\left(\frac{\mu^2}{\alpha h^2}\right)^{1/3},
\end{equation} 
as in \citet{GoldreichSari2003}. Note that $x_{\rm R}<w$ for a gap with a significant depth.

\section{Non-local deposition}\label{sec:non_local}

The assumption of local (instantaneous) wave dissipation and angular momentum deposition breaks down for low-mass planets, on which we focus in this paper.

In Appendix \ref{sec:wave} we show, by generalizing the theory of \citet{GoodmanRafikov2001}, that a density wave that is excited by the planetary torque at a distance $x_0\geq h$ from the planet deposits its angular momentum at a distance
\begin{equation}\label{eq:relation}
x=\max\left[x_0,\left(\frac{\Sigma(x)}{\Sigma(x_0)}\frac{x_0^8h^3}{\mu^2}\right)^{1/5}\right].
\end{equation}
In the case of a uniform density (i.e. before a gap is opened), equation \eqref{eq:relation} reproduces the dissipation location of the dominant wave (excited at $x_0=h$), $x/h=\max[1,(\mu_{\rm th}/\mu)^{2/5}]$, as found by \citet{GoodmanRafikov2001}. Our nominal planet ($\mu=10^{-5}$, $h=3\times 10^{-2}$) is somewhat below the thermal mass $\mu<\mu_{\rm th}\equiv h^3$, implying that the angular momentum deposition is not local.\footnote{Equation \eqref{eq:relation} demonstrates that the deposition is even less local for waves that are excited at $x_0>h$. The decreasing density towards the gap's centre $\Sigma(x_0)<\Sigma(x)$ further enhances the non-locality.}
On the other hand, our nominal planet is above the inertial limit, i.e., the minimum mass required to open a gap fast and deep enough such that the variation in the disc's density profile halts the planet's migration \citep[][see derivation in Appendix \ref{sec:halt}]{Rafikov2002_gap}
\begin{equation}\label{eq:inertial}
	\frac{\mu}{\mu_{\rm th}}>\max\left[h^{5/6},\left(\frac{h}{Q}\right)^{5/13}\right],
\end{equation}
where $Q$ is the Toomre stability parameter of the gas disc, which we assume to be stable  \citep[$Q>1$; see][for discussion]{ChiangLaughlin2013,Schlichting2014}. Lower mass planets migrate rapidly due to the asymmetry between the torques from the outer and inner disc \citep[see, e.g., the review by][and references therein]{KleyNelson2012}. See also \citet{HouriganWard84} and \citet{WardHourigan89} for the original derivation of the ``inertial limit'' for the local-deposition case \citep[in addition, see][for a numerical confirmation]{Li2009,Yu2010,FungChiang2017}. 

In the non-local case, the gap's equilibrium density profile is given by a balance between the torque that each annulus generates due to its interaction with the planet and the viscosity at the location where that torque is deposited in the disc
\begin{equation}\label{eq:non_local_profile}
\nu\frac{\der \Sigma}{\der x}=\frac{\Sigma(x_0)\mu^2}{x_0^4}\frac{\der x_0}{\der x}.
\end{equation} 
Equation \eqref{eq:non_local_profile} is a generalization of equation \eqref{eq:visc} and together with the relation between $x$ and $x_0$, given by equation \eqref{eq:relation}, it defines the density profile.

\subsection{Integration scheme}\label{sec:integration}

We integrate equations \eqref{eq:relation} and \eqref{eq:non_local_profile} to obtain equilibrium gap density profiles that incorporate non-local deposition. In contrast to the local-deposition case, $\der\Sigma/\der x$ in equation \eqref{eq:non_local_profile} depends on $x_0<x$ and on $\Sigma(x_0)$. Therefore, we construct our profiles from the inside out. Since the equations are linear in $\Sigma$, we start\footnote{We assume for simplicity that the torque vanishes for $x_0<h$. In reality, the torque dramatically decreases with decreasing $x_0$ in that region, but the details of the torque density there only smooth the bottom of the gap, without significantly affecting our main results \citep[see][for a similar cutoff]{Kanagawa2015}.} from an arbitrary $\Sigma(h)=\Sigma_0=1$ and normalize the resulting profile to $\Sigma_\infty$. We advance $x_0$ in intervals $\der x_0/h=10^{-3}$ and calculate the gravitational torque that each interval generates $\der T=\Sigma(x_0)\mu^2\der x_0/x_0^4$. This torque raises the profile by $\der\Sigma=\der T/\nu$ over a distance $\der x$ which we find below by advancing $x$. We keep track of the location $x(x_0)$ where the torque is deposited using equation \eqref{eq:relation}. Explicitly, we calculate $\Sigma(x)$ using the following scheme:
\begin{subequations}
	\begin{equation}
	\Sigma\leftarrow\Sigma(x)+\frac{\Sigma(x_0)\mu^2}{x_0^4}\frac{\der x_0}{\nu}
	\end{equation}
	\begin{equation}
	x_0\leftarrow x_0+\der x_0
	\end{equation}
	\begin{equation}
	x\leftarrow\left(\frac{\Sigma}{\Sigma(x_0)}\frac{x_0^8h^3}{\mu^2}\right)^{1/5}
	\end{equation}
	\begin{equation}
	\Sigma(x)\leftarrow \Sigma.
	\end{equation}
\end{subequations}
Finally, we linearly interpolate $\Sigma(x)$ between $x(x_0)$ of consecutive steps.

An example of a density profile that accounts for wave propagation and non-local angular momentum deposition is given in Fig. \ref{fig:compare_local}. As Fig. \ref{fig:compare_local} demonstrates, for our typical parameters, the density at the bottom (centre) of non-local profiles is significantly higher than in the corresponding local-deposition profiles. Due to their milder descent towards the centre, our non-local profiles do not violate the Rayleigh stability criterion (see Section \ref{sec:rayleigh}). 

In Section \ref{sec:understand} we take a closer look at the resulting density profile and explain its shape. A comparison of our scheme with \cite{Duffell2015} is provided in Fig. \ref{fig:duffell}.

\subsection{Understanding the profile}\label{sec:understand}

The shape of the density profile, and specifically its depth and width, can be understood analytically. We start by discussing the waves generated closest to the planet, at $x_0=h$. These waves dissipate and impart their angular momentum at $l_0\equiv h(\mu_{\rm th}/\mu)^{2/5}$, as seen from equation \eqref{eq:relation} and \citet{GoodmanRafikov2001}. Therefore, the density profile is constant for $x<l_0$, and we define $\Sigma_0\equiv\Sigma(x<l_0)$. 

\begin{figure}
	\includegraphics[width=\columnwidth]{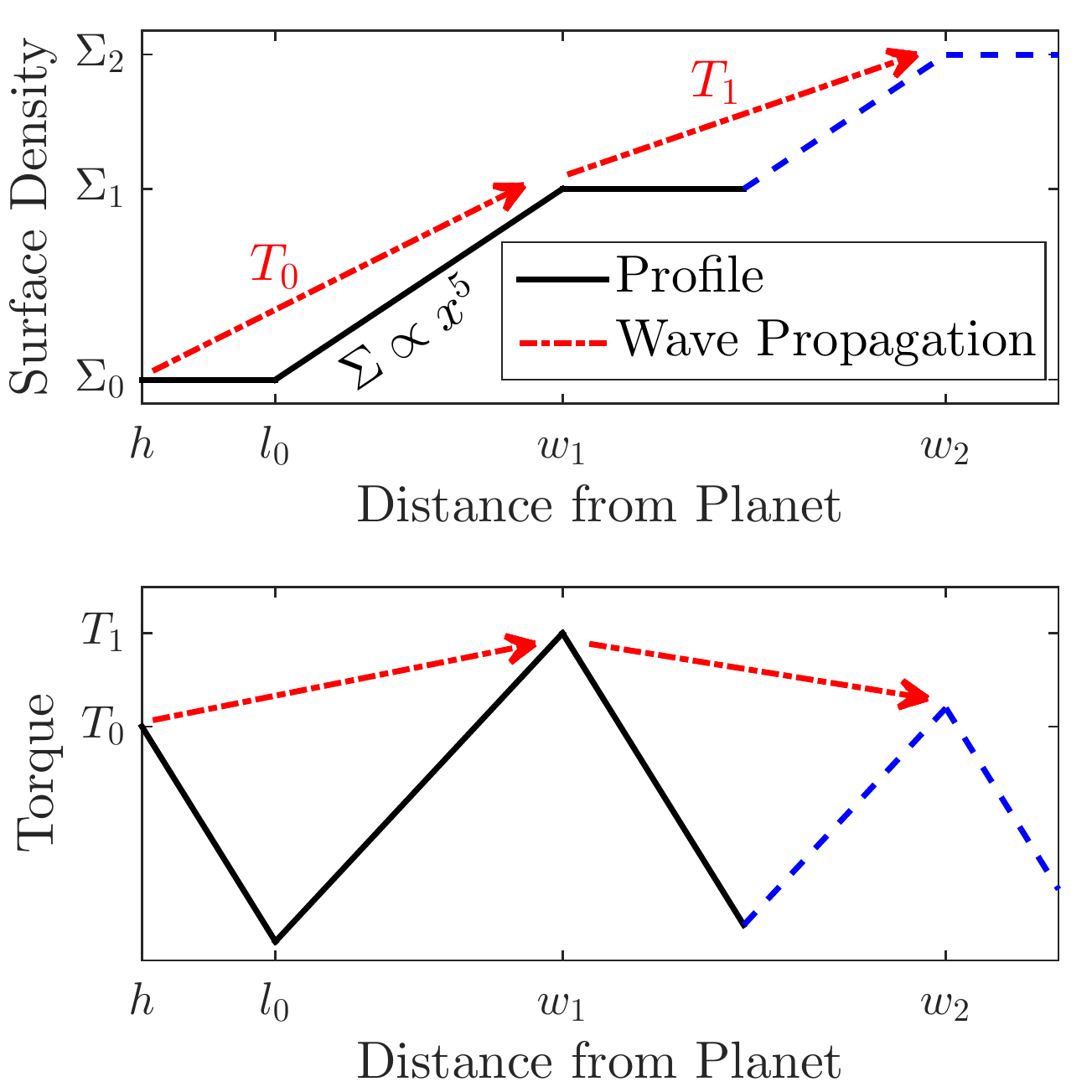}
	\caption{{\it Top panel:} schematic equilibrium gap density profile, normalized to the density at the bottom of the gap $\Sigma_0$. $\Sigma_{1,2}$, $l_0$ and $w_{1,2}$ are given in Section \ref{sec:understand}. The second step (dashed blue) is significant for low values of the viscosity parameter $\alpha$. \newline {\it Bottom panel:} the torque $(x\cdot{\der T}/{\der x}=\Sigma\mu^2/x^3$) generated at each location in the gap. This torque is deposited farther away from the planet (dotted-dashed red arrows). Since $\Sigma_2/\Sigma_1\approx 1+T_1/T_0$ (see Section \ref{sec:understand}), the ratio between the peaks in this panel determines whether the second step in the density profile is significant.}
	\label{fig:schematic}
\end{figure}

We solve equations \eqref{eq:relation} and \eqref{eq:non_local_profile} for the waves that originate from the flat region $h<x_0<l_0$, where $\Sigma(x_0)=\Sigma_0$. While a formal solution is provided in Appendix \ref{sec:flat}, it is more instructive to derive the answer using the schematic Fig. \ref{fig:schematic}. The torque from $h<x_0<l_0$ is dominated by $x_0\sim h$ and it equals $T_0=\Sigma_0\mu^2/h^3$. This torque raises the density profile at $x>l_0$. According to equation \eqref{eq:relation}, since the deposition becomes less local as $\Sigma(x)$ increases, the density profile rises as $\Sigma\propto x^5$ for waves that originate from roughly the same $x_0\sim h$. The rise saturates at $\Sigma_1$ (see Fig. \ref{fig:schematic}) which is found by a torque balance consideration: $\nu\Sigma_1\approx T_0+\nu\Sigma_0$, where $\nu\Sigma_0$ is the viscous torque at $l_0$, $\nu\Sigma_1$ is the viscous torque at $w_1$ (the saturation location), and $T_0$ is the total torque deposited at $l_0<x<w_1$ (it is generated at $x_0\sim h$). The first density step (solid black line) in Fig. \ref{fig:schematic} is thus given by
\begin{equation}\label{eq:sigma1}
\frac{\Sigma_1}{\Sigma_0}=1+\frac{\mu^2}{\nu h^3}=1+\frac{\mu^2}{\alpha h^5}.
\end{equation}
  
Motivated by their two-dimensional simulations (though, for heavier planets and more viscous discs), \citet{Fung2014} derived the same analytical scaling as in equation \eqref{eq:sigma1} for the gap's depletion level \cite[it was also found empirically by][]{DuffellMacFadyen2013}. We emphasize that this scaling assumes that the total torque that the planet exerts is dominated by $x_0\sim h$. It is invalid in the local-deposition case because the integrated torque is dominated by $x>h$ in the steeply ascending density profile. When the non-local deposition is treated properly (and therefore, in the hydrodynamical simulations as well), the density profile rises up much more gently (Fig. \ref{fig:compare_local}), justifying this scaling \citep[see also the discussion in][]{Kanagawa2015}. However, as evident from the bottom panel of Fig. \ref{fig:schematic}, the torque is not always dominated by $x_0=h$ even in the non-local case, necessitating a correction to the scaling of \citet{Fung2014}, which we discuss below.

The first density step saturates $x=w_1$ (see Fig. \ref{fig:schematic}), which can be calculated using equation \eqref{eq:relation} with $x_0\sim h$ and with $\Sigma(w_1)/\Sigma(x_0)=\Sigma_1/\Sigma_0$ given by equation \eqref{eq:sigma1}:
\begin{equation}\label{eq:w1}
w_1=\left(\frac{h^6}{\alpha}\right)^{1/5}.
\end{equation}

As depicted in Fig. \ref{fig:schematic}, the planet excites additional density waves by interacting with the saturated profile. The angular momentum carried by these waves is dominated by the torque $T_1=\Sigma_1\mu^2/w_1^3$ that is excited at $x_0\sim w_1$. These waves induce a second step in the density profile (dashed blue line), which can be studied similarly to the first one. Specifically, by writing a torque balance equation, $\nu\Sigma_1+T_1=\nu\Sigma_2$, we find that 
\begin{equation}\label{eq:sigma2}
\frac{\Sigma_2}{\Sigma_1}=1+\frac{\mu^2}{\nu w_1^3}=1+\frac{\mu^2}{\alpha^{2/5}h^{28/5}}
\end{equation}
and by applying equation \eqref{eq:relation} for $x_0=w_1$ we find that the second step saturates at
\begin{equation}\label{eq:w2}
w_2=\left(\frac{h^7}{\alpha^2}\right)^{1/5}.
\end{equation}

By inspecting equations \eqref{eq:sigma1} and \eqref{eq:sigma2} we deduce that the first step is significant (changes the density by more than a factor of order unity) if $\alpha<\mu^2/h^5\approx 4\times 10^{-3}$ (for our nominal $\mu=10^{-5}$ and $h=3\times 10^{-2}$) whereas the second step is significant if $T_1>T_0$ (see Fig. \ref{fig:schematic}), i.e. $\alpha<\mu^5/h^{14}\approx 2\times 10^{-4}$ (for $\mu<\mu_{\rm th}$, the $\alpha$ required for the second step is always smaller). 

In principle, the pattern of steps in the density $\Sigma$ (top panel of Fig. \ref{fig:schematic}) and peaks in the torque $T$ (bottom panel) continues with
\begin{subequations}\label{eq:n}
	\begin{equation}\label{eq:sigman}
	\frac{\Sigma_{n+1}}{\Sigma_n}=1+\frac{T_n}{\Sigma_n \nu}=1+\frac{\mu^2}{\nu w_n^3}
	\end{equation}
	\begin{equation}\label{eq:wn}
	w_n=w_{n-1}\left(\frac{h}{\alpha}\right)^{1/5}=h\left(\frac{h}{\alpha}\right)^{n/5},
	\end{equation}
\end{subequations}  
with equation \eqref{eq:wn} derived from equations \eqref{eq:relation} and \eqref{eq:sigman} assuming $\mu^2/(\nu w_n^3)\gg 1$ (valid for all the steps of significant depth). However, since $w_n$ increases with $n$ ($\alpha<h$ for $\mu<\mu_{\rm th}$ as long as there is at least one ascent), successive density steps become increasingly less significant according to equation \eqref{eq:n}. In this case ($\mu<\mu_{\rm th}$), the third step is always insignificant, i.e. $\Sigma_3/\Sigma_2\approx 1$, because $\mu^2/(\nu w_2^3)<1$. Therefore, in this paper, we consider only the first two steps ($n=2$), introducing a correction to the simple $n=1$ analytical scaling in previous studies \citep{Fung2014,Duffell2015,Kanagawa2015}. 

\begin{figure}
	\includegraphics[width=\columnwidth]{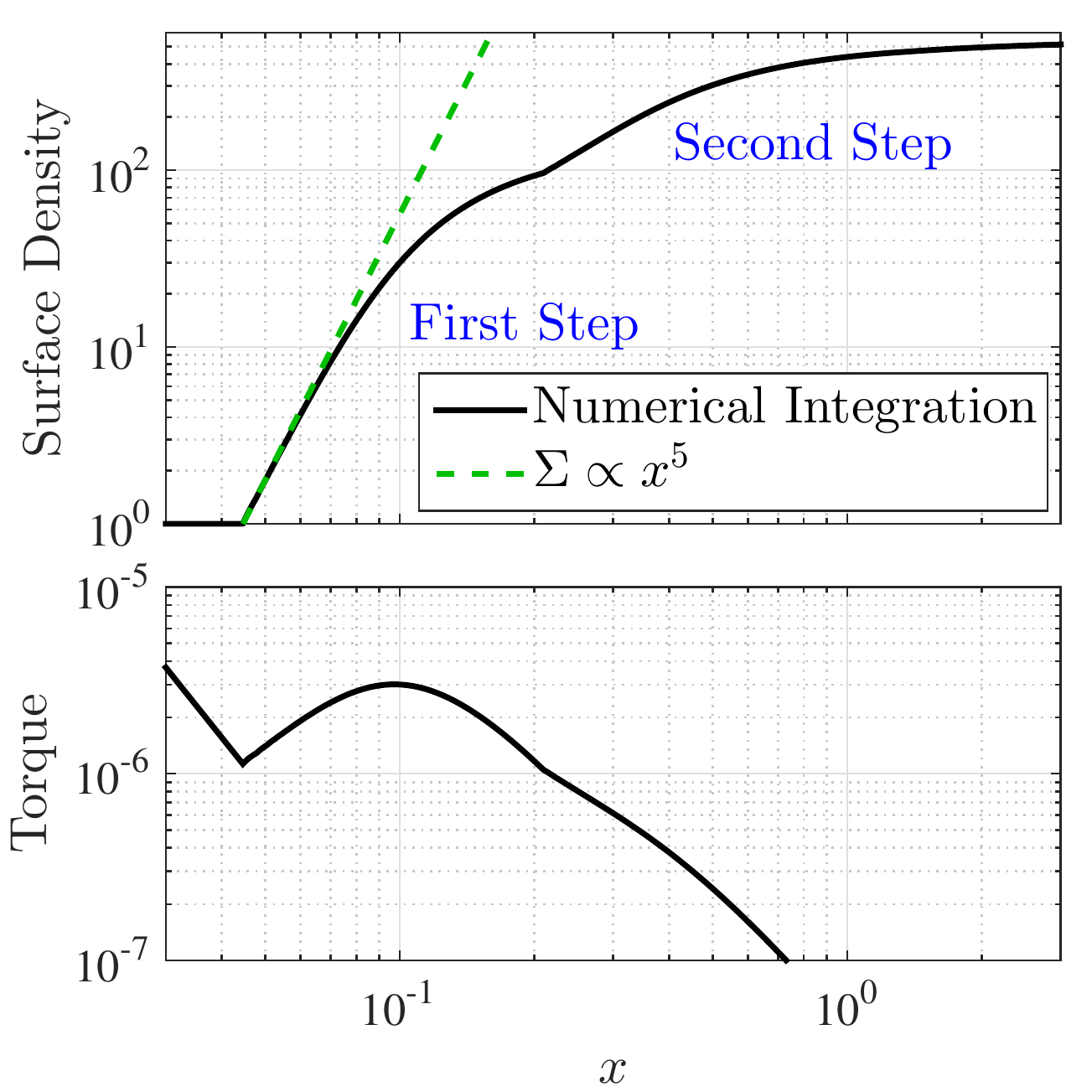}
	\caption{Same as Fig. \ref{fig:schematic}, but calculated using the numerical integration scheme described in Section \ref{sec:integration} for a planet mass $\mu=10^{-5}$, disc scale height $h=3\times 10^{-2}$ and viscosity parameter $\alpha=10^{-5}$ (same as Fig. \ref{fig:compare_local}). The second torque peak is comparable to the first (bottom panel), leading to a factor of $\approx 5$ density increase in the second step (top panel). The third torque peak (bottom panel of Fig. \ref{fig:schematic}) is negligible and appears as an almost unnoticeable bump (bottom panel, at $x>0.2$).}
	\label{fig:calculated}
\end{figure}

In Fig. \ref{fig:calculated} we reinspect the numerically integrated density profile presented in Fig. \ref{fig:compare_local} and compare it to the schematic picture described above. Fig. \ref{fig:calculated} clearly exhibits two peaks in the torque, one at $x=h$ and another at $x=w_1\approx 0.1$, leading to two steps in the density profile. In this case, the two peaks are of comparable magnitude, leading to a factor of a few density increase in the second step, in accordance with equation \eqref{eq:sigma2}. More accurately, the second peak is broader than the first, explaining its significant contribution although it is slightly lower than the first peak. While the first step rises according to the $\Sigma\propto x^5$ analytical result found above, the second step rises more moderately. This is also a consequence of the second peak's breadth, which implies that a range of increasing values of $x_0$ contribute to the rise in $\Sigma$, in contrast to a single $x_0=h$ that is responsible for the first step. Such an increase in $x_0$ generates a shallower rise according to equation \eqref{eq:relation}.

\section{Results}\label{sec:results}

\begin{figure}
	\includegraphics[width=\columnwidth]{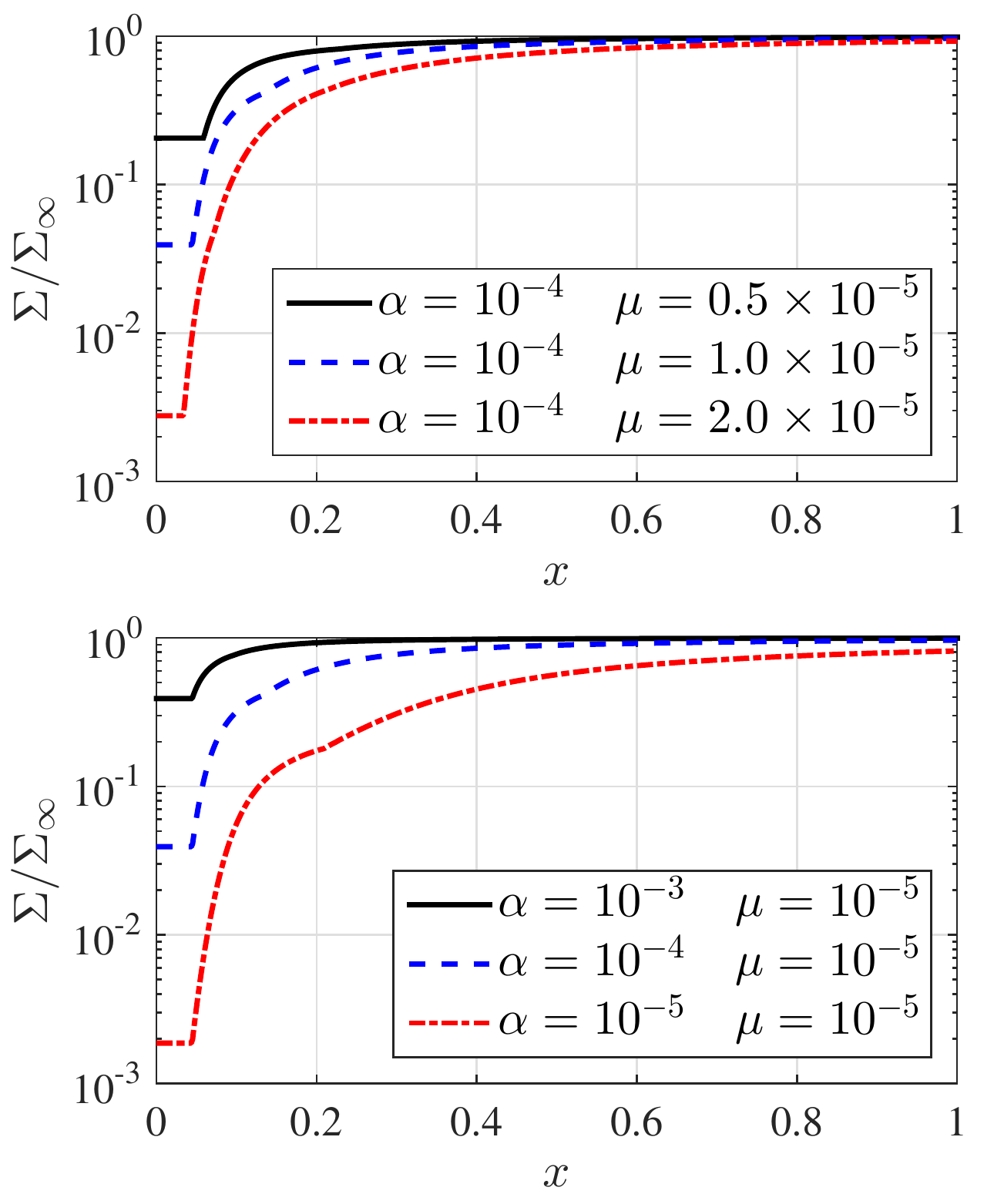}
	\caption{Equilibrium surface density profiles of gaps that are opened by planets with masses $\mu$ in discs with viscosities $\alpha$. The disc scale height is $h=3\times 10^{-2}$, and the non-locality of the angular momentum deposition is accounted for (see Section \ref{sec:non_local}).}
	\label{fig:mu_alpha}
\end{figure}

In Fig. \ref{fig:mu_alpha} we present the density profiles of gaps that are opened by super Earths ($1.7-6.7\,{\rm M}_{\earth}$, assuming a solar mass star) in low-viscosity discs. Intuitively, the gaps grow deeper and wider for more massive planets or less viscous discs (see Sections \ref{sec:depth} and \ref{sec:width} for details).

\subsection{Gap depth}\label{sec:depth}

How depleted are the centres of gaps, or what is $\Sigma_0/\Sigma_\infty$, where $\Sigma_0$ is the density at the centre ($x=0$) and $\Sigma_\infty$ is the unperturbed density? It is evident from Fig. \ref{fig:mu_alpha} that massive super Earths in low-viscosity discs carve out cavities as deep as $\Sigma_0/\Sigma_\infty\sim 10^{-3}$. We discuss some implications of this depletion in Section \ref{sec:accretion}.

\begin{figure}
	\includegraphics[width=\columnwidth]{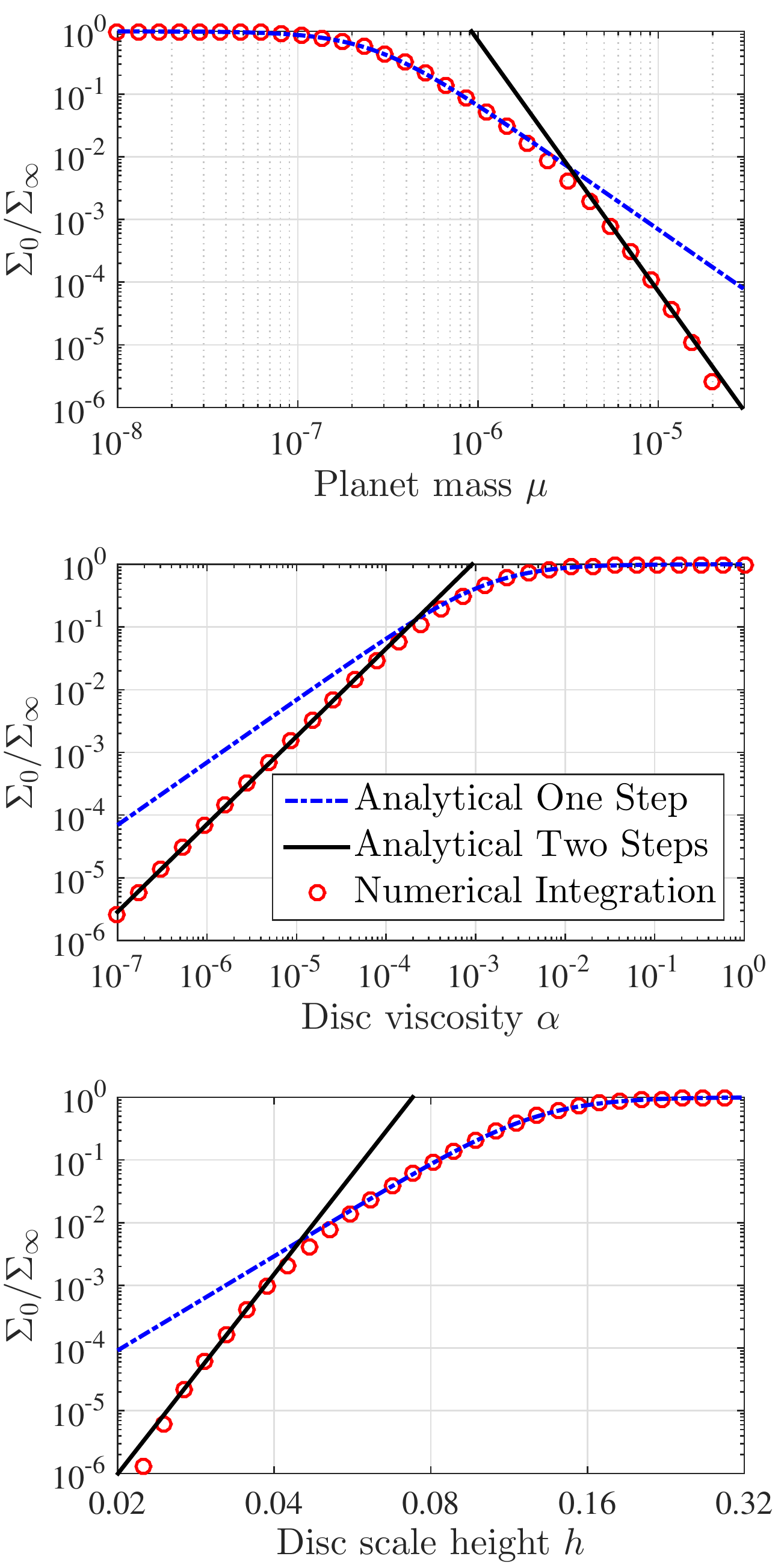}
	\caption{Depletion level, i.e., the ratio of the density at the bottom of the gap $\Sigma(x=0)$ to the unperturbed density $\Sigma(x\to\infty)$, as a function of the planet's to star's mass ratio ($\mu$, top panel), the disc's viscosity parameter ($\alpha$, middle panel), and the disc's aspect ratio ($h$, bottom panel). The nominal values are $\mu=10^{-5}$, $\alpha=10^{-6}$, and $h=3\times 10^{-2}$. Our results (red circles) are calculated using an integration scheme that is described in Section \ref{sec:integration}. The one-step analytical formula (dotted-dashed blue lines), originally from \citet{DuffellMacFadyen2013} and \citet{Fung2014}, is given by equation \eqref{eq:sigma1}. The second step correction (solid black lines), $\Sigma_0/\Sigma_\infty=\alpha^{7/5}h^{53/5}\mu^{-4}$, is taken from equation \eqref{eq:depth}. A coefficient $\approx 2.5$ was added to the analytical expressions to fit the numerical results.}
	\label{fig:formula}
\end{figure}

In Fig. \ref{fig:formula} we present a systematic study of the gap's depth as a function of our three parameters: $\mu,
\alpha,h$. The results are explained well by our analytical theory (Section \ref{sec:understand}). Explicitly, we combine equations \eqref{eq:sigma1} and \eqref{eq:sigma2} and derive the following expression for the gap's depletion:

\begin{equation}\label{eq:depth}
\frac{\Sigma_0}{\Sigma_\infty}\approx\begin{cases}
\begin{aligned}
& 1 & \displaystyle{\alpha>\frac{\mu^2}{h^5}=h\left(\frac{\mu}{\mu_{\rm th}}\right)^2} \quad &\textrm{(no gap)} \\
& \displaystyle{\frac{\alpha h^5}{\mu^2}} &\displaystyle{\frac{\mu^5}{h^{14}}<\alpha<\frac{\mu^2}{h^5}} \quad &\textrm{(one step)} \\
& \displaystyle{\frac{\alpha^{7/5}h^{53/5}}{\mu^4}} &\displaystyle{\alpha<\frac{\mu^5}{h^{14}}=h\left(\frac{\mu}{\mu_{\rm th}}\right)^5} \quad &\textrm{(two steps)}.
\end{aligned}
\end{cases}
\end{equation}
Equations \eqref{eq:sigma1} and \eqref{eq:sigma2} also prescribe the interpolation between the different regimes of equation \eqref{eq:depth}.

\begin{figure}
	\includegraphics[width=\columnwidth]{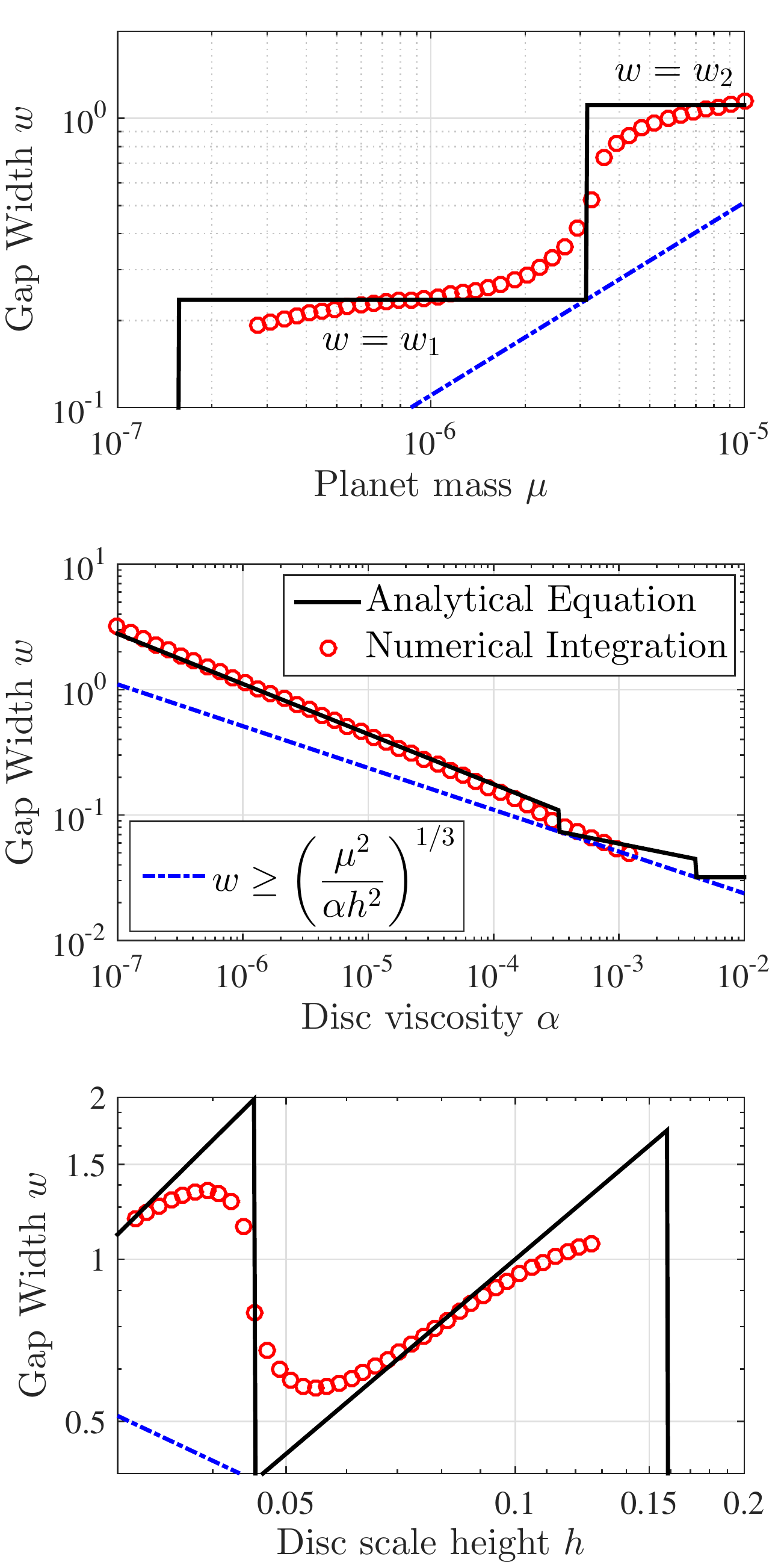}
	\caption{The gap's width, defined as $x$ for which $\Sigma(x)/\Sigma_\infty=1/2$, as a function of the planet's to star's mass ratio ($\mu$, top panel), the disc's viscosity parameter ($\alpha$, middle panel), and the disc's aspect ratio ($h$, bottom panel). The nominal values are $\mu=10^{-5}$, $\alpha=10^{-6}$, and $h=3\times 10^{-2}$. Our results (red circles) are calculated using an integration scheme that is described in Section \ref{sec:integration}. Note that the width is ill-defined for gaps that are too shallow. The analytical solution (solid black lines) is according to equation \eqref{eq:width}, where $w=w_0=h$ was assumed for shallow gaps ($\alpha>\mu^2/h^5$). A lower limit $w\gtrsim (\mu^2/\alpha h^2)^{1/3}$, which is explained in Section \ref{sec:width}, is also plotted (dotted-dashed blue lines). A coefficient $\approx 1.6$ was added to the analytical expressions to fit the numerical results.}
	\label{fig:width_curve}
\end{figure}

Fig. \ref{fig:formula} and equation \eqref{eq:depth} demonstrate that the standard analytical scaling \citep{DuffellMacFadyen2013,Fung2014,Duffell2015,Kanagawa2015}, which assumes a single step, describes well gaps that are not too deep, yet it is invalid in the low-viscosity regime ($\alpha\lesssim 10^{-4}$ for our nominal $\mu$ and $h$), where a second step in the density profile further depletes the gap with respect to the surrounding disc. The standard expression underestimates the gap's depth by up to two orders of magnitude in the surveyed parameter range. According to equation \eqref{eq:depth}, the transition between the two regimes is when $\Sigma_0/\Sigma_\infty\sim (\mu/\mu_{\rm th})^3$. The numerical results in Fig. \ref{fig:formula} deviate from the analytical two-step scaling for high masses ($\mu\gtrsim 3\times 10^{-5}$) or low scale heights ($h\lesssim 0.02$) because our assumption of $\mu<\mu_{\rm th}\equiv h^3$ breaks down (see Section \ref{sec:thermal_mass}). In addition, we expect some inaccuracy in our model for $\alpha\lesssim 3\times 10^{-7}$ (with the nominal $h$) due to the excitation of extremely low-order resonances (see Section \ref{sec:wide_gaps} for an estimate of these deviations and for a discussion of the range of validity of our model).

\subsection{Gap width}\label{sec:width}

Another interesting quantity is the gap's width, defined here as $x$ for which $\Sigma(x)/\Sigma_\infty=1/2$. Analytically, the width $w$ is calculated in Section \ref{sec:understand}, specifically equations \eqref{eq:w1} and \eqref{eq:w2}:
\begin{equation}\label{eq:width}
\frac{w}{a}\equiv w=\begin{cases}
\begin{aligned}
& w_1=\displaystyle{\left(\frac{h^6}{\alpha}\right)^{1/5}} &\displaystyle{\frac{\mu^5}{h^{14}}<\alpha<\frac{\mu^2}{h^5}} \quad &\textrm{(one step)} \\
& w_2=\displaystyle{\left(\frac{h^7}{\alpha^2}\right)^{1/5}} &\displaystyle{\alpha<\frac{\mu^5}{h^{14}}} \quad &\textrm{(two steps)}.
\end{aligned}
\end{cases}
\end{equation}
For insignificant gaps ($\alpha>\mu^2/h^5$), the width is ill-defined since the density never drops below $\Sigma_\infty/2$.

In Fig. \ref{fig:width_curve} we present the widths of numerically integrated gap profiles as a function of $\mu,\alpha,h$. The non-trivial dependence of the width $w$ on $\mu$ and $h$ is explained well by the analytical equation \eqref{eq:width}. In particular, a jump of order $(h/\alpha)^{1/5}=\mu_{\rm th}/\mu$ at the transition from a one-step to a two-step density profile $\mu=(\alpha h^{14})^{1/5}$, as well as the non-monotonicity of $w(h)$ are reproduced. The non-monotonicity is a result of a drop in $w$ in the transition from a two-step (low $h$) to a one-step (high $h$) regime, whereas $w(h)$ increases in each of the regimes.
Quantitatively, the width of our widest gaps is comparable to the separation of the planet from the star (i.e. $w\sim 1$). While this is one of our important results, it is also one of our sources of inaccuracy (see Section \ref{sec:wide_gaps}). 

It is noteworthy that equation \eqref{eq:sigman} provides a lower limit on the width $w$. If the profile has $n$ density steps, by definition $\Sigma_{n+1}/\Sigma_n\sim 1$, leading to $\mu^2/(\nu w_n^3)\lesssim 1$. In other words, the gap's width satisfies
\begin{equation}\label{eq:width_boundary}
w=w_n\geq\left(\frac{\mu^2}{\alpha h^2}\right)^{1/3}.
\end{equation} 
By comparing equations \eqref{eq:width_local} and \eqref{eq:width_boundary} we find that gaps are always wider than predicted by the local-deposition assumption, except for the transitions between regimes, in which case both predictions are similar (see Fig. \ref{fig:width_curve}).  

\subsection{Time to reach equilibrium}\label{sec:time}

In equilibrium, the gravitational torque is balanced by the viscosity. Therefore, the time it takes to open a gap equals to the viscous time-scale to close it:
\begin{equation}\label{eq:time_n}
t=\frac{w_n^2}{\nu}=\frac{1}{\alpha}\left(\frac{h}{\alpha}\right)^{2n/5},
\end{equation}
where we substitute the gap's width from equation \eqref{eq:wn}. Explicitly, for $n=1,2$ steps:
\begin{equation}\label{eq:time}
\Omega t\equiv t=\begin{cases}
\begin{aligned}
& \displaystyle{\left(\frac{h^2}{\alpha^7}\right)^{1/5}} &\displaystyle{\frac{\mu^5}{h^{14}}<\alpha<\frac{\mu^2}{h^5}} \quad &\textrm{(one step)} \\
& \displaystyle{\left(\frac{h^4}{\alpha^9}\right)^{1/5}} &\displaystyle{\alpha<\frac{\mu^5}{h^{14}}} \quad &\textrm{(two steps)},
\end{aligned}
\end{cases}
\end{equation}
exhibiting a jump of order $(\mu_{\rm th}/\mu)^2$ at the transition, following the jump in $w$.

Equation \eqref{eq:time} demonstrates that the equilibrium time increases rapidly with decreasing viscosity, especially in the low-viscosity (two-step) regime. This is one of the major obstacles for full hydrodynamical simulations of the problem. Quantitatively, since $t$ is the number of orbits required to reach equilibrium in our normalized units (see Section \ref{sec:local}), we notice that it becomes comparable to the gas disc's lifetime \citep[a few million years, or $\sim 10^8$ orbits for our 10 day orbit; see][]{Mamajek2009,WilliamsCieza2011,Alexander2014} for $\alpha\lesssim 10^{-5}$. If the gas disc's lifetime is determined by its viscous evolution then the time to open a gap of width $w\sim 1$ is, by definition, comparable to that lifetime. In addition, while type I migration is cancelled out by the disc's feedback, as discussed in Section \ref{sec:non_local} and Appendix \ref{sec:halt}, the planet might still experience type II migration \citep{LinPapaloizou86,Ward97}. In the classical picture, the rate of this migration is dictated by the disc's viscous timescale, which is comparable, as explained above, to the gap opening time of our widest gaps ($w\sim 1$). This might introduce order of unity corrections for such gaps, whereas narrower gaps are opened much faster than the planet travels across them.

\section{Summary and discussion}\label{sec:summary}

The density profiles of gaps that planets open in protoplanetary gas discs have been the focus of intense research in the past decade. In principle, the equilibrium gap profile can be calculated analytically by balancing the planet's gravitational torque, which opens the gap, with the disc's viscosity, which suppresses it \citep{TanigawaIkoma07}. In practice, however, the torque generated by low-mass planets is carried away by density waves and deposited in the disc only where these waves steepen into a shock \citep{GoodmanRafikov2001}. This non-local angular momentum deposition complicates the balance calculation.

Here, we adapted the wave propagation mechanism of \citet{GoodmanRafikov2001} to the case where a deep gap has already formed and derived an analytical relation between the wave excitation and deposition locations. We then incorporated this relation into a simple one-dimensional integration and obtained density profiles that properly account for the non-local deposition. Our method enables us to easily probe lower disc viscosities in comparison with previous multidimensional full hydrodynamical simulations that suffer from numerical viscosity and take a long time to converge to a steady state (see Section \ref{sec:time}).

We complemented the numerical integration with an analytical understanding of the gap's profile (Section \ref{sec:understand}). While previous analytical calculations \citep[motivated by hydrodynamical simulations of more viscous discs; see][]{DuffellMacFadyen2013,Fung2014} assumed that the gravitational torque is dominated by the interaction of the planet with the gas annulus at a distance $h$ (the disc's scale height) from it, we found that for low viscosities, $\alpha<h(\mu/\mu_{\rm th})^5$ ($\approx 10^{-4}$ for our nominal planet), the torque is dominated by a second peak, forming a two-step density profile (see Figs \ref{fig:schematic} and \ref{fig:calculated}). We derived equations \eqref{eq:depth} and \eqref{eq:width} for the depth and width of such two-step profiles and demonstrated that the standard one-step formula, which has been applied in previous studies, underestimates the depletion (depth) of gaps in low-viscosity discs (Fig. \ref{fig:formula}). In addition, we demonstrated that two-step gaps can be as wide as the planet's separation from the star (Fig. \ref{fig:width_curve}). Finally, we estimated the time to reach a steady state in equation \eqref{eq:time}, providing a reference to check the convergence of future hydrodynamical simulations. 

\subsection{Caveats and approximations}\label{sec:caveats}

In this section we discuss the validity of our approximations and remaining caveats.

\subsubsection{Rossby wave instability}

The gap profiles that we studied here are gradual enough such that their curvature does not violate the Rayleigh stability criterion (see Section \ref{sec:rayleigh}). However, multidimensional instabilities, such as the Rossby wave instability \citep{Li2000}, were not considered in this work. These instabilities might limit the gap depletion.

Similarly to \citet{Kanagawa2017}, we extrapolate the scaling of \citet[][see their Table 2]{Ono2016} to the wide ($w\sim 1$, see Fig. \ref{fig:width_curve}) gaps in the two-step regime and estimate that the Rossby wave instability might limit the depletion of these gaps to $\sim 10^{-2}$. However, \citet{Ono2016} studied narrower ($w\leq 0.2$) Gaussian gaps in thicker ($h\geq 0.1$) discs. It is yet unclear whether their results can be extrapolated to gaps of a different shape and parameter range. Moreover, \citet{Ono2016} find a considerable sensitivity to the equation of state of the gas, indicating (by extrapolating their Figure 13 and Table 2) that wide isothermal gaps can grow orders of magnitude deeper than adiabatic ones, without triggering an instability.

In the future, the \citet{Lovelace1999} criterion can be applied to our gaps and discs in order to check whether they are susceptible to the Rossby wave instability. Such a test, which is similar to Fig. 12 in \citet{Kanagawa2017}, is beyond the scope of our one-dimensional model.

\subsubsection{Wide gaps}\label{sec:wide_gaps}

Our widest gaps are comparable in size to the planet's orbital separation (the approximation $x
\ll 1$ breaks down; see Figs \ref{fig:mu_alpha} and \ref{fig:width_curve}). Such gaps, in which the disc's scale height and unperturbed density also vary, should be treated more carefully in the future \citep{Rafikov2002}.
The problem is less severe than one might expect because while the width of two-step gaps is given by $w_2\sim 1$, the gravitational torque that forms them is excited at $w_1\ll w_2$ (see Section \ref{sec:understand}).
	
\begin{figure}
	\includegraphics[width=\columnwidth]{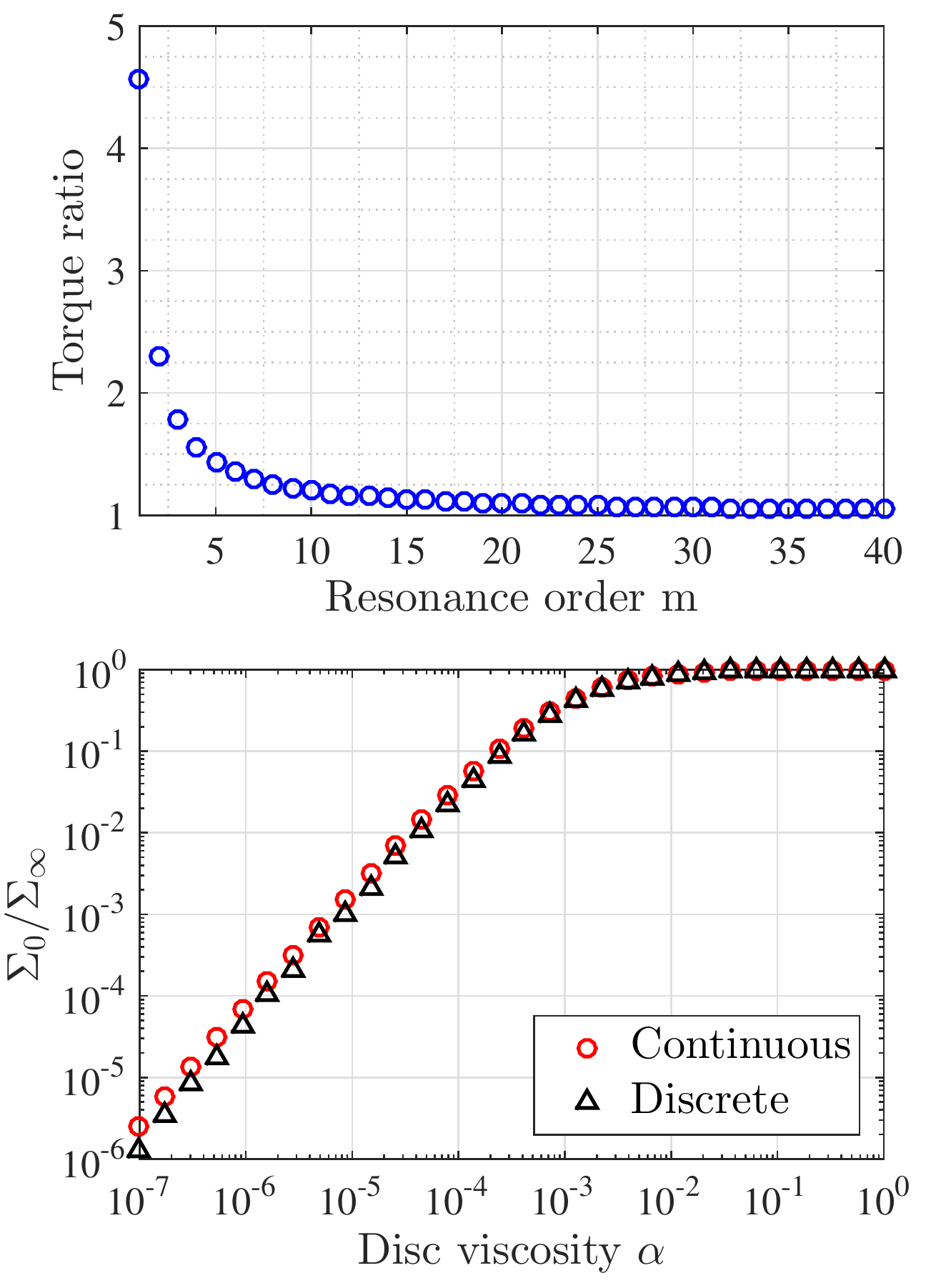}
	\caption{{\it Top panel:} The ratio of the accurate (discrete) torque, as a function of the outer Lindblad resonance order $m$, to the ${\der T}/{\der x}\propto x^{-4}$ approximation for $m\gg 1$ ($x\ll 1$). The accurate $m$th order torque $T_m$ is given by equations (7)--(9) in \citet{Ward97}, where we have approximated $mh\ll 1$, since the torque cutoff at $x<h$ is already accounted for in our integration scheme (we are interested here in the deviation at $m\sim 1$). The continuous (approximate) power-law torque density is integrated between adjacent resonances to produce the cumulative torque, which is then compared to $T_m$.
    \newline {\it Bottom panel:} Gap depth as a function of the disc's viscosity parameter $\alpha$ for $\mu=10^{-5}$ and $h=3\times 10^{-2}$, calculated using our standard approximate continuous scheme (red circles, same as in Fig. \ref{fig:formula}) and using a discrete scheme which excites density waves only at outer Lindblad resonances (black triangles). The discrete torques $T_m$ are accurately calculated according to \citet{Ward97}, as in the top panel (with the correct coefficients to match the schemes for $m\gg 1$).}
	\label{fig:discrete}
\end{figure}

A particular concern is that, in the two-step regime, the planet's torque is dominated by interaction with gas that is relatively far away from the planet ($x=w_1$, see Fig. \ref{fig:schematic}), whereas previous studies considered torques that are generated at a distance $h\ll 1$. This implies that the torque in our case is dominated by relatively low-order (low $m$) Lindblad resonances, which deviate from the standard high-$m$ approximation, given by equation \eqref{eq:torque}. In the top panel of Fig. \ref{fig:discrete} we present the ratio of the accurate (discrete) torque, as calculated by \citet{Ward97}, to the $m\gg 1$ approximation ${\der T}/{\der x}\propto x^{-4}$ that we use (see the caption of Fig. \ref{fig:discrete} for details). Quantitatively, the second torque peak in our nominal model is located at $x\approx 0.1$ (Fig. \ref{fig:calculated}), which corresponds to the $m\approx 7$ outer and inner Lindblad resonances. According to Fig. \ref{fig:discrete} (top panel), our high-$m$ approximation underestimates the torque strength for this $m$ by about 30 percent (in the outer disc; in the inner one, it slightly overestimates). In the bottom panel of Fig. \ref{fig:discrete} we use a discrete version of our numerical integration scheme (Section \ref{sec:integration}) which excites density waves only at outer Lindblad resonance locations (we calculate only the outer half of the gap). Fig. \ref{fig:discrete} demonstrates that a discrete scheme, which also incorporates the exact torque strength $T_m$ (see top panel) leads to slightly deeper gaps compared to our standard continuous scheme, without modifying our main conclusions. The deviation increases with decreasing $\alpha$, as lower-$m$ resonances become important, because $w_1\propto \alpha^{-1/5}$ according to equation \eqref{eq:w1}.

We note that our wave dissipation formula is still given by the generalization of the \citet{GoodmanRafikov2001} results, even in the discrete scheme. We expect deviations from these results for waves generated at low-$m$ resonances, because the shearing-sheet approximation (i.e. $x\ll 1$) breaks down, and the shape of the generated wave might be different. Nonetheless, we expect such corrections to be of order unity at most, since $w_1\approx 0.1\ll 1$ for our nominal parameters (or equivalently $m\approx 7$ is high enough), leading to mild modifications, similarly to Fig. \ref{fig:discrete}. The situation might be different though for extremely low viscosities $\alpha\sim 10^{-7}$, for which lower $m\sim 1$ resonances become important.

Quantitatively, we estimate that our model is accurate as long as $w_1=(h^6/\alpha)^{1/5}\lesssim 0.3$. In this range, the wave is excited at resonances $m>2$, for which the discrete nature of the torque introduces a correction of less than a factor of 2, according to Fig. \ref{fig:discrete}. In this range we also expect that the shearing-sheet approximation for the excited torque is reasonable. For our nominal $h=3\times 10^{-2}$, our model is therefore valid for $\alpha>3\times 10^{-7}$. For $\alpha=10^{-6}$, which is used in Fig. \ref{fig:formula}, $w_1\lesssim 0.3$ for $h\lesssim 0.04$. For $h>0.04$, however, the gap is in the one-step regime (see Fig. \ref{fig:formula}), so the torque is excited at $h\ll 1$, ensuring that our model is accurate for the entire presented $h$ range.

\subsubsection{Thermal mass non-linearity}\label{sec:thermal_mass}	

In this paper we focused on planets below the thermal mass, i.e. $\mu<\mu_{\rm th}\equiv h^3$. Above the thermal mass, the density waves are highly non-linear and they shock as soon as they form, modifying our scaling relations \cite[see Section \ref{sec:non_local} and][]{GoodmanRafikov2001}. Because our nominal planets are only a factor of a few lighter than the thermal mass, it is worthwhile to check whether this non-linearity affects our results.

\begin{figure}
	\includegraphics[width=\columnwidth]{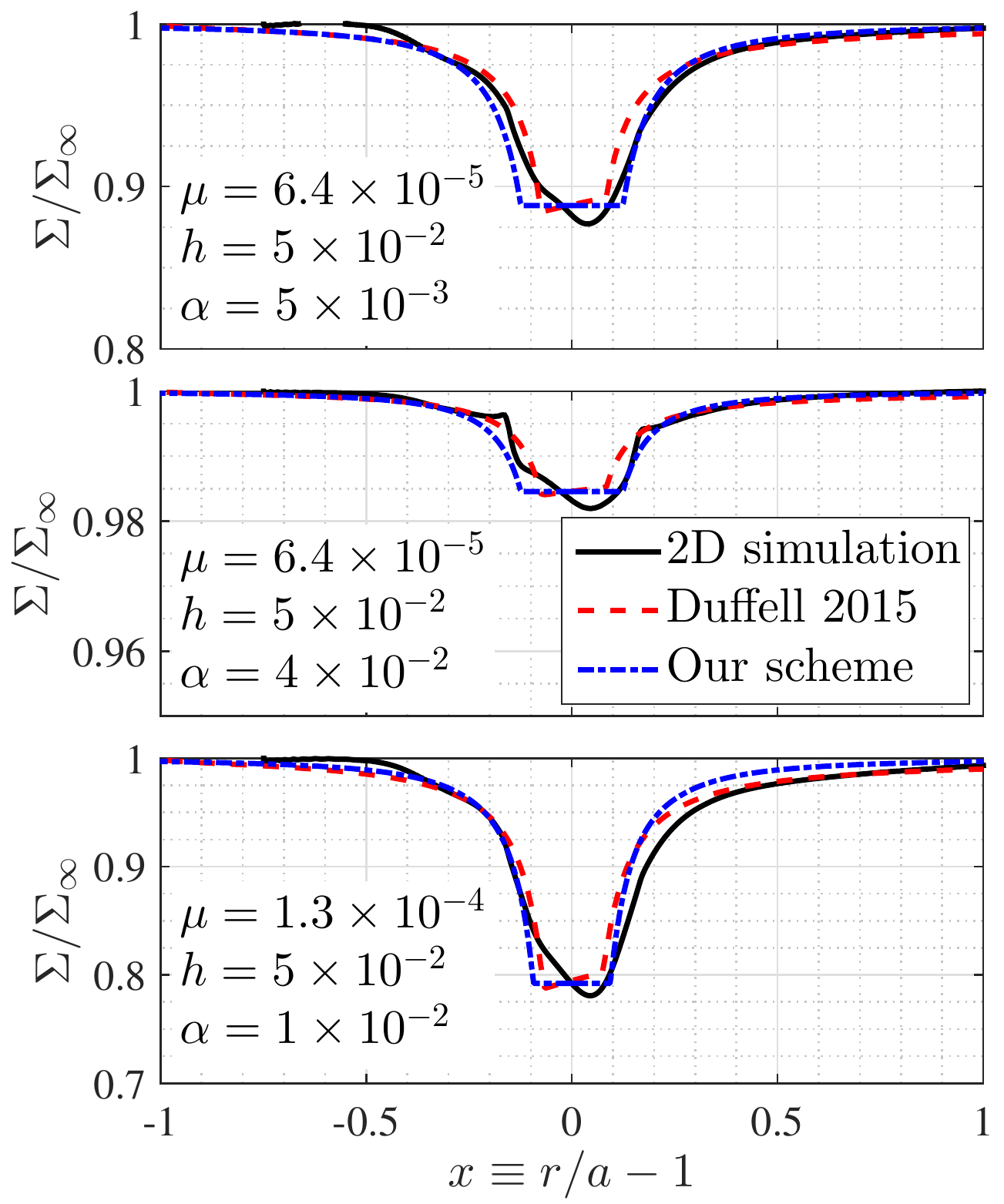}
	\caption{Comparison of shallow gap profiles that are calculated using our scheme (dotted-dashed blue lines) to 2D hydrodynamical simulations (solid black lines) and to the analytical model (dashed red lines) of \citet[][the solid black lines are also taken from that paper]{Duffell2015}. For consistency between the schemes, we reintroduced order of unity coefficients for the torque strength \citep[$f_0/\upi$ with $f_0=0.45$; see][]{Duffell2015} and for the shocking distance \citep[1.9, which is where two thirds of the angular momentum are deposited in the disc according to][this is the only free calibration parameter in the figure]{GoodmanRafikov2001}. Note the different scale of the vertical axes.}
	\label{fig:duffell}
\end{figure}

In Fig. \ref{fig:duffell} we compare our calculations to 2D hydrodynamical simulations that were presented by \citet{Duffell2015}. We reproduce three representative profiles in the regime $\mu/\mu_{\rm th}\lesssim 1$ \cite[for higher masses, both our model, and the analytical model of][are invalid]{Duffell2015}. Our profiles fit the full hydrodynamical calculations reasonably well for both $\mu/\mu_{\rm th}=0.5$ (top two panels) and $\mu/\mu_{\rm th}=1.0$ (bottom panel), affirming the validity of the $\mu<\mu_{\rm th}$ approximation, at least for shallow gaps. 

However, when a deep gap is opened by a massive planet ($\mu\gtrsim\mu_{\rm th}$), non-linear effects can also drive higher-order (secondary, tertiary, etc.) spiral arms at distant locations from the planet. These higher-order waves, which we do not calculate in this paper, can interfere with the angular momentum transport far from the planet. They can even shock and carve out secondary gaps \citep[see, e.g.,][and references therein]{ArtymowiczLubow92,FungDong2015,Juhasz2015,Lee2016spiral,Bae2017,Kanagawa2017}. Whether or not these higher-order spirals affect the wide gaps that lower-mass planets ($\mu\lesssim\mu_{\rm th}$) open in low-viscosity discs remains to be tested in multidimensional hydrodynamical simulations. In this context, it is noteworthy that \citet{Bae2017} find that secondary spirals can carve out gaps even by planets well below the thermal mass (at least in the inner disc) if the disc's viscosity is low enough. Since gaps grow wider with decreasing $\alpha$ (see Section \ref{sec:width}), it is likely that these secondary gaps merge with the primary gap for sufficiently low viscosities.
	 	
\subsubsection{Wave decay and interference}

	We assumed, for simplicity, that a wave that is excited at $x_0$ deposits all its angular momentum at the shocking distance $x(x_0)$, which is given by equation \eqref{eq:relation}. But, according to \citet{GoodmanRafikov2001}, a fraction of the angular momentum is deposited farther away from the planet.	Taking this into account might alter the shape of our density profiles \citep[see, however,][who find only a weak dependence on the smearing width]{Kanagawa2015}.
	
	Following the assumption above, and since $w_1\gg h $, we treat the waves that originate from $x\sim h$ and those from $x\sim w_1$ (i.e., the two torque peaks; see Fig. \ref{fig:schematic}) independently. More generally, we assumed that the waves that originate from each annulus evolve independently. While the calculation of the first step's depth ($\Sigma_1/\Sigma_0$) does not rely on this assumption (because it is given solely by $T_0$), its width ($w_1$) might be sensitive to the interference of different waves. This, in turn, might introduce a correction to $T_1=\Sigma_1\mu^2/w_1^3$, and thereby alter the height of the second step $\Sigma_2/\Sigma_1$.

\subsection{Atmosphere accretion}\label{sec:accretion}

Part of our motivation for studying gap depths is the accretion of gas atmospheres onto rocky cores. Such gas envelopes, of a few percent in mass, are a natural explanation to the ubiquitous low-density super Earths in the {\it Kepler} sample \citep[e.g.][]{Lopez2012,Lissauer2013}. In principle, the gas accretion rate, and therefore the mass of the atmosphere, depends on the gas density inside the gap that surrounds the planet. In practice, however, this dependence is logarithmic \citep{Rafikov2006,PisoYoudin2014,Ginzburg2016,LeeChiang2016}. Therefore, only very deep gaps, such as the ones we find here for low-viscosity discs, are relevant. Quantitatively, by examining equations (13) and (16) of \citet{Ginzburg2016} and Figure 4 of \citet{LeeChiang2016}, we estimate that $\Sigma_0/\Sigma_\infty$ must be as low as $10^{-2}-10^{-3}$ in order to reduce the atmosphere mass by a factor of 2. Fig. \ref{fig:formula} shows that super Earths on orbits of $\sim$ 10 days can carve out such deep gaps if the \citet{ShakuraSunyaev73} viscosity parameter is $\alpha\lesssim 10^{-5}$. As discussed in Section \ref{sec:time}, the time it takes to excavate such a gap is comparable to the disc's lifetime.

We conclude that super Earths in low-viscosity discs starve by opening deep gaps around themselves. The gaps may explain why these planets did not acquire more gas and did not grow into Jupiters via runaway gas accretion \citep{Lee2014}. While there are other explanations \citep[][]{InamdarSchlichting2015,Ormel2015,LeeChiang2016,GinzburgSari2017} to the sub runaway (lighter than the core) mass of super Earth atmospheres, we find the starvation by gaps appealing, because it implies that planets regulate their own atmosphere accretion, without relying on external processes.

\section*{Acknowledgements}

This research was partially supported by ISF (Israel Science Foundation) and iCore (Israeli Centers of Research Excellence) grants. SG thanks Eugene Chiang, Paul Duffell, Jeffrey Fung, and Eve Lee for warm hospitality at UC Berkeley and for discussions that initiated this work. We also thank Paul Duffell for comments on the paper's draft and for providing us with the curves from his 2015 paper. Finally, we thank Kazuhiro D. Kanagawa for a helpful review that improved the paper. 




\bibliographystyle{mnras}
\bibliography{gap} 



\appendix
\section{Wave Propagation}\label{sec:wave}

\citet{GoodmanRafikov2001} and \citet{Rafikov2002} formally derived the radial distance from the planet where density waves dissipate, by reducing the problem to Burgers' equation. Here, we outline the main concepts behind that calculation and derive the more general equation \eqref{eq:relation}.

We consider a spiral (due to the differential rotation) density wave that is excited at a distance $x_0\geq h$ from the planet. Since the velocity of a sound wave depends on its amplitude (i.e. density), it will steepen into a shock after travelling a radial distance of $x=\lambda\Sigma/\Delta\Sigma$, with $\lambda=x_0h/x$ denoting the radial wavelength \citep[given by the azimuthal wavelength $x_0$ times the pitch angle of the spiral; see][]{GoldreichTremaine78,Rafikov2002} and $\Delta\Sigma$ the perturbation in density ($\Sigma$ is the unperturbed surface density). 

The amplitude of the wave $\Delta\Sigma(x)$ can be found by considering the angular momentum flux $F$, which is conserved until the wave shocks (since no angular momentum is imparted to the disc prior to dissipation). Each harmonic $m$ (with an amplitude $\Delta\Sigma_m\equiv\Sigma_m-\Sigma$) carries an angular momentum flux
\begin{equation}\label{eq:f_m}
F_m=\int{\Sigma_m v_x v_\theta{\rm d}\theta}=x_0e \Delta\Sigma_m=\lambda x_0\frac{\Delta\Sigma_m^2}{\Sigma},
\end{equation}
where $v_x=e$ is the radial velocity (or eccentricity) of the perturbed gas and $v_\theta$ is the azimuthal velocity. The perturbation in density is related to the eccentricity by $\Delta\Sigma_m/\Sigma=e/\lambda$ (intuitively, consider the separation change between two particles which are separated by $\lambda/2$) and the azimuthal width of the wave is $\sim x_0$. Note that the first-order term in equation \eqref{eq:f_m} $v_x\Sigma$ averages out in each wavelength, so the flux is quadratic $F_m\propto v_x\Delta\Sigma_m\propto\Delta\Sigma_m^2$ \citep{LandauLifshitz}. Also note that we substitute $v_\theta=1$ because $\Delta v_\theta\sim e^2/x$ \citep[e.g.][]{LubowIda2010} introduces a higher-order effect ($\Delta v_\theta \Sigma < \Delta\Sigma_m$).

We calculate the total angular momentum flux by summing over all the harmonics and using Perseval's theorem \citep{GoldreichTremaine80,Rafikov2002}
\begin{equation}\label{eq:f_tot}
F=\sum_m{F_m}=\frac{\lambda x_0}{\Sigma}\int{\Delta\Sigma^2{\rm d}\theta}=\lambda x_0^2\Sigma\left(\frac{\Delta\Sigma}{\Sigma}\right)^2.
\end{equation}
Finally, we equate $F$ to the total torque generated by the gas at $x_0$, $T=\Sigma(x_0)\mu^2/x_0^3$, which is given by integration of equation \eqref{eq:torque}, and obtain the amplitude of the wave
\begin{equation}
\left(\frac{\Delta\Sigma}{\Sigma}\right)^2=\frac{\Sigma(x_0)}{\Sigma(x)}\frac{\mu^2}{\lambda x_0^5}.
\end{equation}
By substituting $\lambda$ and after some algebra, we extract the shocking distance and arrive at equation \eqref{eq:relation}. 

\section{Dissipation of waves generated by a flat profile}\label{sec:flat}

In this section we solve equations \eqref{eq:relation} and \eqref{eq:non_local_profile} for the special case where $\Sigma(x_0)=\Sigma_0=1$, e.g., for the waves generated in the flat region $h<x_0<l_0$ (as explained in Section \ref{sec:integration}, we can normalize the density profile to $\Sigma_0$ which is chosen as unity to simplify the equations). The resulting density profile is explained more intuitively in Section \ref{sec:understand}, and the following explicit calculation is given for completeness.

By substituting $\Sigma(x_0)=1$, $\nu=\alpha h^2$, and combining equations \eqref{eq:relation} and \eqref{eq:non_local_profile} we find
\begin{equation}\label{eq:deriv_flat}
\frac{\der \Sigma}{\der x}=\frac{5\beta x^{-23/8}\Sigma^{3/8}}{1+\beta x^{-15/8}\Sigma^{-5/8}},
\end{equation}
where $\beta\equiv(1/8)\mu^{5/4}h^{-7/8}\alpha^{-1}$. We integrate equation \eqref{eq:deriv_flat} from $\Sigma(l_0)=1$ (as discussed in Section \ref{sec:understand}, the waves that are excited at $x_0=h$ dissipate at $x=l_0$) by distinguishing the two different regimes of the denominator 
\begin{equation}\label{eq:flat_exact}
\frac{\Sigma(x)}{\Sigma_1}\approx\begin{cases}
\displaystyle{\left(\frac{x}{w_1}\right)^5} & l_0<x<w_1\\
\displaystyle{\left(1+A\left[1-\left(\frac{x}{w_1}\right)^{-15/8}\right]\right)^{8/5}} & x>w_1,
\end{cases}
\end{equation}
where $\Sigma_1/\Sigma_0\sim \mu^2/(\alpha h^5)$, $w_1\sim (h^6/\alpha)^{1/5}$, and $A>0$ is a coefficient of order unity.

The conclusion from equation \eqref{eq:flat_exact} is that the flat density profile at the bottom of the gap generates waves that raise the gap's profile as $\Sigma\propto x^5$ until $x\sim w_1$, where the density saturates at $\Sigma\sim\Sigma_1$ (the density only increases by an order of unity factor at $x>w_1$). This result is also explained in Section \ref{sec:understand}.

\section{Halting migration by disc feedback}\label{sec:halt}

\cite{Rafikov2002_gap} generalized previous studies \citep{HouriganWard84,WardHourigan89} and derived the ``inertial limit'', which is given by equation \eqref{eq:inertial}. Planets above this limit are massive enough to change the disc's density profile, leading to a termination of type I migration. In this section we briefly outline the derivation.

In an unperturbed disc, planets migrate due to an asymmetry of order $h$ between the inner and outer one-sided torques, which leads to a two-sided net torque \cite[see][and references therein]{KleyNelson2012}
\begin{equation}
T_{2s}=hT_{1s}=\frac{\Sigma\mu^2}{h^2}.
\end{equation}
As discussed in Section \ref{sec:non_local}, the planet tries to open a gap which is initially of width $l_0\equiv h(\mu_{\rm th}/\mu)^{2/5}$, where density waves dissipate. The time it takes a planet to migrate over such a gap is given by
\begin{equation}
t_{\rm mig}=\frac{\mu l_0}{T_{2s}},
\end{equation}
and the change in the angular momentum of the gas $\Delta\Sigma l_0^2$ at $x=l_0$ is found by multiplying this time by the one-sided torque. The resulting change in the density at $x=l_0$ is given by
\begin{equation}\label{eq:inertial_delta_sigma}
\Delta\Sigma=\frac{\mu}{l_0h}.
\end{equation}

Using equation \eqref{eq:torque}, the pile up of gas at $x=l_0$ changes the torque on the planet by $\Delta\Sigma\mu^2/l_0^3$. This is enough to balance $T_{2s}$ and halt the planet's type I migration if
\begin{equation}\label{eq:inertial_frac}
\frac{\Delta\Sigma}{\Sigma}>\frac{l_0^3}{h^2}.
\end{equation}

Equation \eqref{eq:inertial_frac} prescribes the conditions for halting the planet's migration by this disc feedback. For any disc feedback to be able to halt the planet, we require
\begin{equation}\label{eq:interial1}
\frac{l_0^3}{h^2}<1 \Rightarrow \frac{\mu}{\mu_{\rm th}}>h^{5/6}.
\end{equation}
For the density perturbation generated during the planet's migration to suffice, we substitute $\Delta\Sigma$ from equation \eqref{eq:inertial_delta_sigma} into equation \eqref{eq:inertial_frac}:
\begin{equation}\label{eq:interial2}
\frac{\mu}{\mu_{\rm th}}>\Sigma^{5/13}=\left(\frac{h}{Q}\right)^{5/13},
\end{equation} 
where $Q\sim h/\Sigma$ is Toomre's stability parameter in our normalized units.

By combining equations \eqref{eq:interial1} and \eqref{eq:interial2} we derive equation \eqref{eq:inertial}, which reproduces equations (41) and (53) of \citet{Rafikov2002_gap}. Note that this condition is more strict than equation (54) of \citet{Rafikov2002_gap} because \citet{Rafikov2002_gap} is also interested in planets that open deep gaps (i.e. $\Delta\Sigma/\Sigma>1$) without halting their migration. By substituting $\Delta\Sigma$ from equation \eqref{eq:inertial_delta_sigma}, this occurs if $\mu/\mu_{\rm th}>Q^{-5/7}$. We, on the other hand, are interested only in (almost) static planets that have enough time to fully open their deep and wide gaps, until equilibrium is reached (see Section \ref{sec:time}).


\bsp	
\label{lastpage}
\end{document}